\documentclass[twocolumn,pra,nobalancelastpage,aps,raggedbottom, superscriptaddress]{revtex4-2}
\usepackage{times}

\usepackage{amsmath,amsfonts,amssymb,graphicx,hyperref,epstopdf,xcolor,tikz,scalerel,natbib,array}
\usepackage{mathptmx,textcomp,braket}
\usepackage[T1]{fontenc}
\usepackage[utf8]{inputenc}
\usepackage{multirow}
\usepackage{array}

\definecolor{lime}{HTML}{A6CE39}
\DeclareRobustCommand{\orcidicon}
{
	\begin{tikzpicture} 
	\draw[lime, fill=lime] (0,0) circle [radius=0.15] node[white] {{\fontfamily{qag}\selectfont \tiny ID}};
	\draw[white, fill=white] (-0.0625,0.095) 	circle [radius=0.007];
	\end{tikzpicture}
	\hspace{-2.2mm}
}
\newcommand\orcidID[1]{\href{https://orcid.org/#1}{\orcidicon}}

\newcommand{\be}{\begin {equation}}
\newcommand{\ee}{\end {equation}}
\newcommand{\beqa}{\begin {eqnarray}}
\newcommand{\eeqa}{\end {eqnarray}}
\newcommand{\mb}{\mathbf}

\hypersetup{colorlinks,citecolor=blue,filecolor=black,linkcolor=blue,urlcolor=blue}

\begin{document}

\title{Tailoring polarisation of attosecond pulses via co-rotating bicircular laser fields}

\author{Rambabu Rajpoot\orcidID{0000-0002-2196-6133}}
\email[E-mail: ]{ramrajpoot3@gmail.com}
\affiliation{Department of Physics, Birla Institute of Technology and Science - Pilani, Rajasthan, 333031, India.}

\author{Amol R. Holkundkar\orcidID{0000-0003-3889-0910}}
\email[E-mail: ]{amol@holkundkar.in}
\affiliation{Department of Physics, Birla Institute of Technology and Science - Pilani, Rajasthan, 333031, India.}

\author{Navdeep Rana}
\affiliation{Department of Physics, Indian Institute of Technology Bombay, Powai, Mumbai, 400076, India.}

\author{Gopal Dixit}
\affiliation{Department of Physics, Indian Institute of Technology Bombay, Powai, Mumbai, 400076, India.}

\date{\today}

\begin{abstract}
The present work introduces  a robust way to generate 
attosecond pulses with tunable ellipticity via high-order harmonic generation 
by co-rotating   $\omega - 2\omega$ bicircular laser fields. 
The total electric field of the laser fields exhibits an absence of rotational symmetry, which leads to the generation of high harmonics of the same helicity across a broad range of spectral bandwidth. 
High-harmonics with the same helicity 
offer the opportunity to synthesize attosecond pulses with tunable ellipticity.   
The polarisation properties of the generated harmonics are robust against the 
variations in driving fields' parameters, such as  
wavelength, intensity ratio, and the sub-cycle phase between $\omega-2\omega$ fields. 
Our work opens an avenue to study chiral-sensitive light-matter ultrafast processes on their intrinsic timescale. 
 
\end{abstract}

\maketitle

\section{Introduction}
High-harmonic generation (HHG) is one of the non-perturbative 
nonlinear processes of laser-matter interaction. 
HHG has been intensively used for a tabletop coherent light source in extreme ultraviolet 
and soft x-ray energy regimes with attosecond temporal resolution~\cite{midorikawa2022progress}. 
HHG in a gaseous medium  proceeds via a three-step process ~\cite{PhysRevLett.70.1599,Corkum1993_PRL}, wherein an intense laser pulse librates  an electron via tunnel ionisation as a first step. 
The liberated electron gains energy in the presence of driving laser as it propagates in the continuum and, eventually, is driven back to recollide with the parent ion. The kinetic energy acquired by the electron is emitted in the form of higher-order harmonics of the driving laser field. 
HHG not only plays a paramount role in producing attosecond pulses, 
but also  offers a wide array of applications 
by unraveling  ultrafast electron dynamics in matter with atomic-scale spatialtemporal resolutions~\cite{Krausz2009_RMP, Corkum2007_NatPhy, Chini2014_nat, Baykusheva2016_PRL, Reich2016_PRL, bredtmann2014x, dixit2012imaging}. 

Owing to the rapid development on both the theoretical and experimental fronts,  researchers have 
focused on controlling the polarization of the emitted harmonics. 
Usually, the polarization of the emitted harmonics is controlled by utilizing various forms of the  counter-rotating  $\omega-2\omega$ bicircular fields configuration, 
such as fields having different intensity ratios, relative phase, and ellipticities \cite{Neufeld2018_PRL, Dorney2017_PRL, Long1995_PRA, Eichmann1995_PRA, Frolov2018_PRL, Rajpoot2021_JPhysB, Milosevic2000_PRA, fleischer2014_NatPhoton, kfir2015_NatPhoton, Milosevic2020_PRA}, non-collinear mixing of combining pulses \cite{Hickstein2015_NatPhoton, Huang2018_NatPhoton}, adding a seed pulse \cite{Dixit2018_PRA}, as well as utilizing the plasmonic field enhancement \cite{Ansari2021_PRA}, to name a few. 
However, due to the three-fold symmetry restriction enforced by the counter-rotating $\omega-2\omega$ fields, harmonics with alternating helicity are generated. 
Such circularly polarized harmonics with alternating helicity could generate linearly polarized attosecond pulses with each subsequent pulse rotated by 120$^{\circ}$  in space \cite{Milosevic2000_PhysRevA}. 
Thus, it is crucial to desire harmonics of the same helicity across a range of spectral bandwidths to produce attosecond pulse with tunable ellipticity.

In this work, we introduce a robust scheme  to generate attosecond pulse with tunable ellipticity using  co-rotating $\omega-2\omega$ circularly polarized  fields. 
In the following, we will demonstrate that the harmonics with the same helicity are produced owing  
to the absence of rotational symmetry in the driving  co-rotating $\omega-2\omega$ bicircular fields. 
The robustness of  our scheme is tested with respect to the variations in the driving  fields' parameters.
It is found that highly elliptical attosecond pulse can be generated as the scheme is 
insensitive to any  variations in the parameters. 
The generated attosecond pulses with circular or elliptical polarization are desirable to 
probe various chiral-sensitive light-matter phenomena \cite{Fan2015_PNAS, Cireasa2015_NatPhys, Tang2011_Scienece, Hendry2010_NatuteNanotech, Baykusheva2018_PRX, giri2021imaging, giri2020time}.
Recently, Lu and co-workers have applied corotating bicircular field configurations with 1:3 ratio to discuss the role of Coriolis-force effect in the generation of high harmonics~\cite{Li2022_PRA}. 
Moreover, the superiority of the molecular target over atomic in the context of corotating bicircular fields
setup is discussed in Ref.~\cite{Qiao2019_PRA}.   
Solids and plasma targets, apart from gaseous systems,  have also attracted attentions for  
HHG via co-rotating driving fields
~\cite{Rana2022_PhysRevAppl, Chen2022_sym, Jia2018_JChemPhys}.

The paper is organized as follows. Details of the numerical methods are discussed in Sec. \ref{sec2}, followed by the results and discussion in Sec. \ref{sec3}. The concluding remarks and future directions are discussed in Sec. \ref{sec4}. 
 
\section{Numerical Methods}
\label{sec2} 
\begin{figure}[t]
	\centering\includegraphics[width=1\columnwidth]{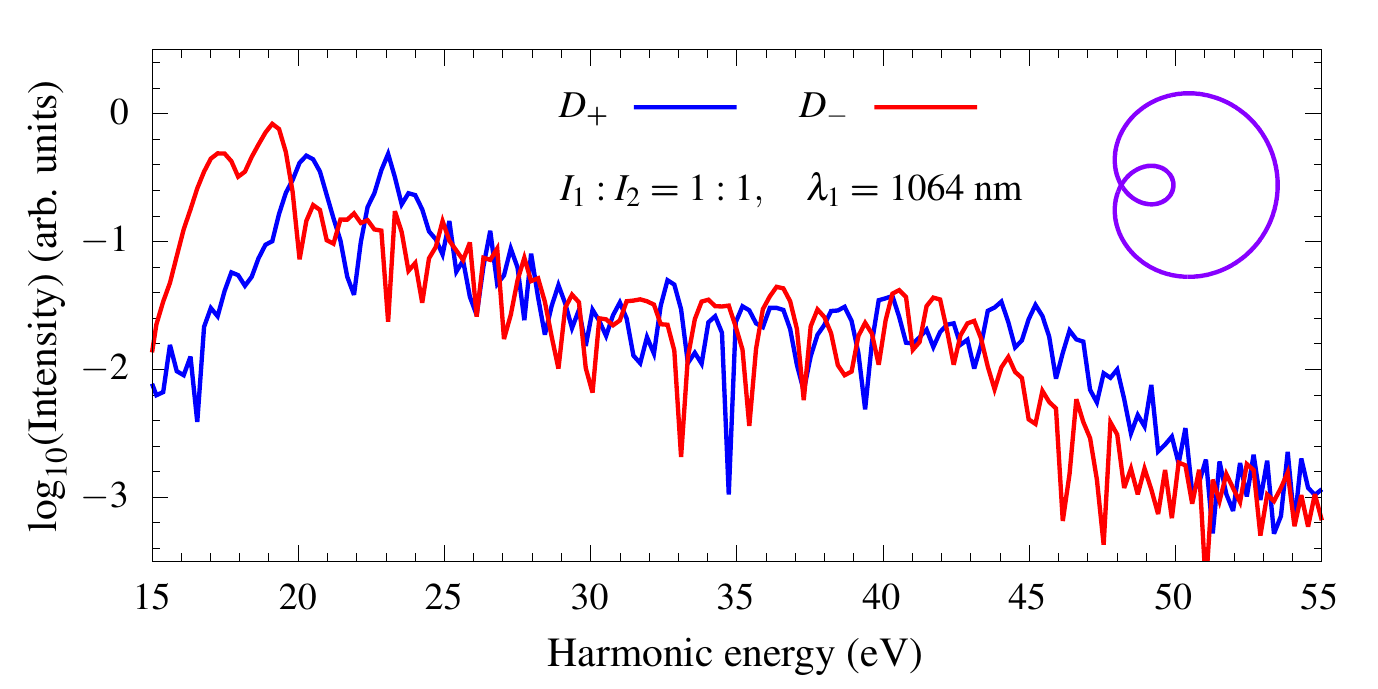}
	\caption{High-harmonic spectrum of helium driven by co-rotating $\omega-2\omega$  fields. 
		$D_{+}$ harmonic component (blue line) is co-rotating with the $\omega$ field, whereas 
		the red line represents the counter-rotating $D_{-}$ harmonic component.. 
		Inset shows Lissajous figure corresponding to the total co-rotating $\omega-2\omega$ electric fields for one cycle of $\omega_1$ field. $\lambda_{1}$ = 1064 nm, $I_{1} = I_{2} = 5\times 10^{13}$ W/cm$^2$, and $\phi = 0^{\circ}$  are used to simulate the spectrum.}
	\label{1064_laser_hhg}
\end{figure}

Time-dependent Schr\"odinger equation in two-dimensions, within single-active-electron approximation, for helium is numerically solved as discussed in Refs.~\cite{Rajpoot2021_JPhysB, Rajpoot_2020}. 
The harmonic  spectrum is obtained by performing the Fourier transform of the dipole acceleration as
\be
S_{\kappa}(\Omega)  = \Big| \frac{1}{\sqrt{2\pi}} \int a_{\kappa}(t) e^{-i\Omega t} dt \Big|^2 = \big| a_{\kappa}(\Omega) \big|^2,
\ee 
where $\kappa$ stands for the $x$ or $y$ components of the time-dependent dipole acceleration.
To describe the polarization of the harmonics,  
the intensity of the left- and right-rotating components is  obtained as 
$ D_{\pm} = \big| a_{\pm}(\Omega) \big|^2$ with $a_{\pm}(\Omega) =  [ a_x(\Omega) \pm i a_y(\Omega) ]/\sqrt{2}$.	
Ellipticity of the emitted harmonics is calculated using
\be
\epsilon = \frac{ |a_+(\Omega)| - |a_-(\Omega)| }{|a_+(\Omega)| + |a_-(\Omega)|}.
\label{ellipticiy}
\ee
The parameter $\epsilon$ varies in the interval from $-1$ to $+1$, and the sign of $\epsilon$ defines the helicity of the harmonics. The harmonics rotating in a counter-clockwise direction have positive helicity while those rotating in a clockwise direction have negative helicity~\cite{Zhang2017_OptLett, Huo2021_PRA}. 

The temporal profile of an attosecond pulse is 
constructed by filtering the desired frequency range with an appropriate window function $w(\Omega)$ and then performing  an inverse Fourier transform as \cite{Rajpoot2023_JPhysB}:
\be \mathcal{E}_{\kappa}(t) = \frac{1}{\sqrt{2\pi}} \int a_{\kappa}(\Omega) w(\Omega) e^{i\Omega t} d\Omega.\ee
Here, $w(\Omega) = \Theta(\Omega - \Omega_1) \Theta(\Omega_2 - \Omega)$, wherein $\Omega_1 \leq \Omega \leq \Omega_2$ is the frequency range to be filtered, and $\Theta(x)$ is standard step function. 

The total electric field corresponding to the $\omega-2\omega$ co-rotating configuration is written as
\be 
\begin{split}
	\mb{E}(t) = f(t)\big\{ &E_1\ [ \cos(\omega_1 t) \mb{\hat{e}}_x + \sin(\omega_1 t) \mb{\hat{e}}_y ] +\\
	&E_2\ [ \cos(\omega_2 t + \phi)  \mb{\hat{e}}_x + \sin(\omega_2 t + \phi) \mb{\hat{e}}_y]\big\}, 
\end{split}
\label{electric_driver}
\ee
where $\omega_j$ and $E_j$ are the frequency and the electric field amplitude of the $j^{\textrm{th}}$ component of the bicircular field, respectively.  $\phi$ defines the sub-cycle phase between the two fields. 
The temporal pulse envelope $f(t) = \sin^2(\pi t/\tau)$ with total duration $\tau = 5T_1$, where $T_1 = 2\pi/\omega_1$ is the period of the fundamental field.

We have considered the spatial simulation domain of $\pm150$ a.u. along both $x$ and $y$ directions. The value of parameter $r_\text{abs} = \pm 142$ a.u. is considered. The spatial step $\Delta x = \Delta y \approx 0.29$ a.u. is used and the simulation time step $\Delta t = 0.01$ a.u. is considered, which is well within the criteria $\Delta t \lesssim 0.5(\Delta x)^2$. The convergence is tested with respect to the spatial grid as well as space and time steps. Our simulation utilizes widely used \textit{Armadillo} library for linear algebra purpose \cite{Sanderson2016}.

In the following sections, we discuss the harmonic generation by co-rotating bicircular laser pulses and the production of highly elliptically polarized attosecond pulses.

\section{Results and Discussions}
\label{sec3}
\begin{figure}[t]
	\centering\includegraphics[width=1\columnwidth]{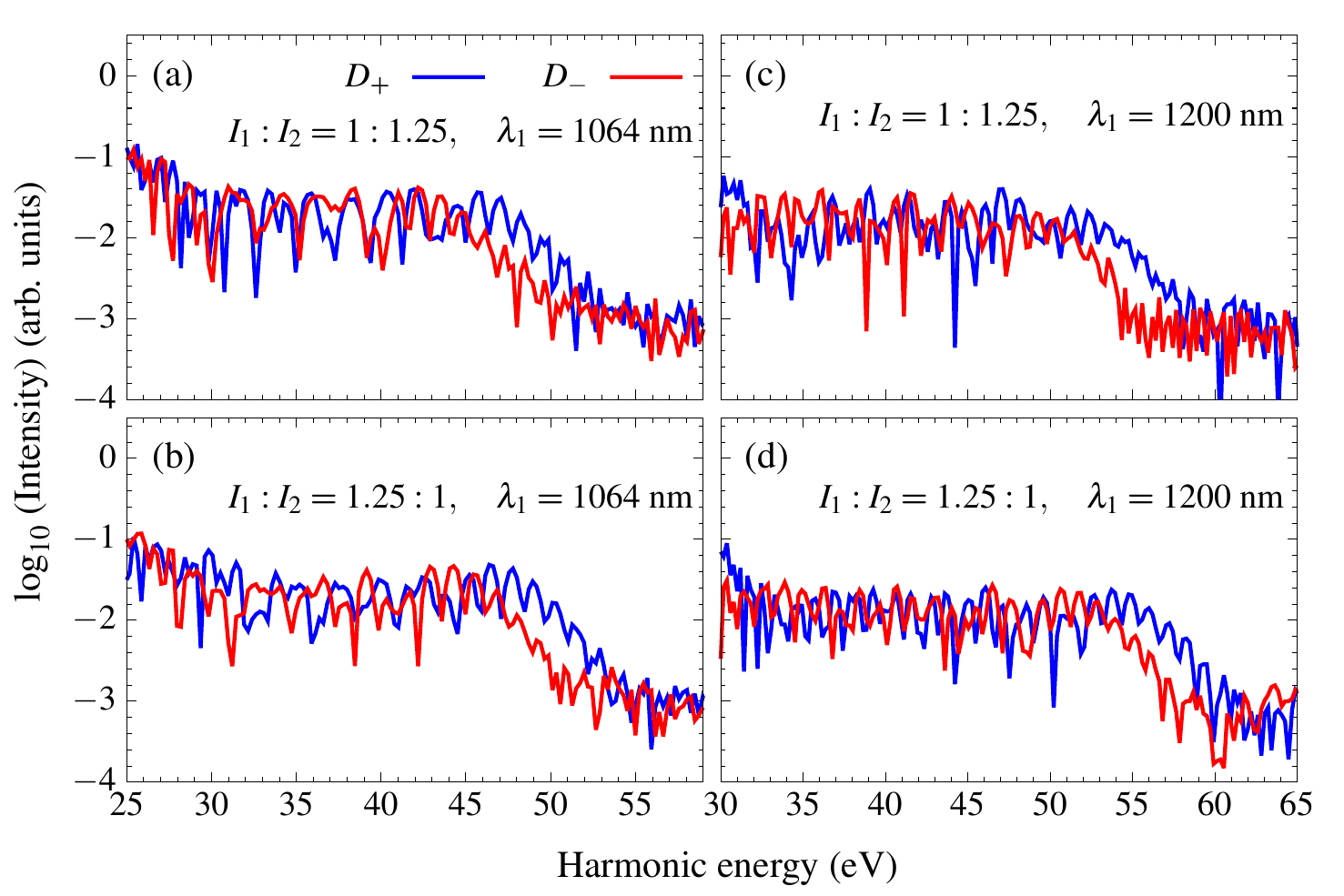}
	\caption{Same as Fig.~\ref{1064_laser_hhg} with different wavelengths, and intensity ratios.  
		The wavelength  is considered $1064$ nm in (a-b) and $1200$ nm in (c-d). 
		Here, value $1$ corresponds to the intensity $5\times 10^{13}$ W/cm$^2$. }
	\label{hhg_1064_1200}
\end{figure} 
The harmonic spectra of helium driven by  co-rotating $\omega-2\omega$ fields is presented in 
Fig. \ref{1064_laser_hhg}. Absence of the dynamical rotational symmetry of the total electric field results in
the generation of even and odd-order harmonics with a regular plateau structure as evident from the figure. 
Additionally, the harmonic component co-rotating with the driving field has  
higher contrast, dominates in the cutoff region.
The present observations are consistent with the propensity rules as discussed in Ref.[~\cite{Alon1998_PRL}]. 
This is in contrast with counter-rotating $\omega-2\omega$ configuration, which results in harmonic doublets with alternating helicity. 

Before we discuss the generation of attosecond pulses with tunable ellipticity from the spectrum shown 
in Fig. \ref{1064_laser_hhg}, let us explore the robustness of the features in the spectrum with respect to 
various laser parameters.  
Figure \ref{hhg_1064_1200} presents harmonic spectra corresponding to co-rotating $\omega-2\omega$ configuration having different fundamental wavelengths and intensity ratios. 
The overall nature of the spectrum remains insensitive with respect to the changes in the intensity ratio as the ratio is tuned from 1:1 to 1:1.25 and 1.25:1, as evident from Figs. \ref{hhg_1064_1200}(a) and \ref{hhg_1064_1200}(b), respectively. 
A similar observation can be made when the wavelength is increased from 1064 nm to 1200 nm with the same intensity ratios [see Figs. \ref{hhg_1064_1200}(c) and \ref{hhg_1064_1200}(d)]. 
Thus, the higher yield of the co-rotating harmonic component in blue is insensitive with respect to the variations in the wavelength as well as  intensity ratio of the co-rotating fields. 
In all cases, the subcycle phase between $\omega-2\omega$ fields,  $\phi$, is zero. 
At this juncture, it is natural to envision how different values of $\phi$ affect the nature of the harmonic spectra discussed so far.

\begin{figure}[t]
	\centering\includegraphics[width=1\columnwidth]{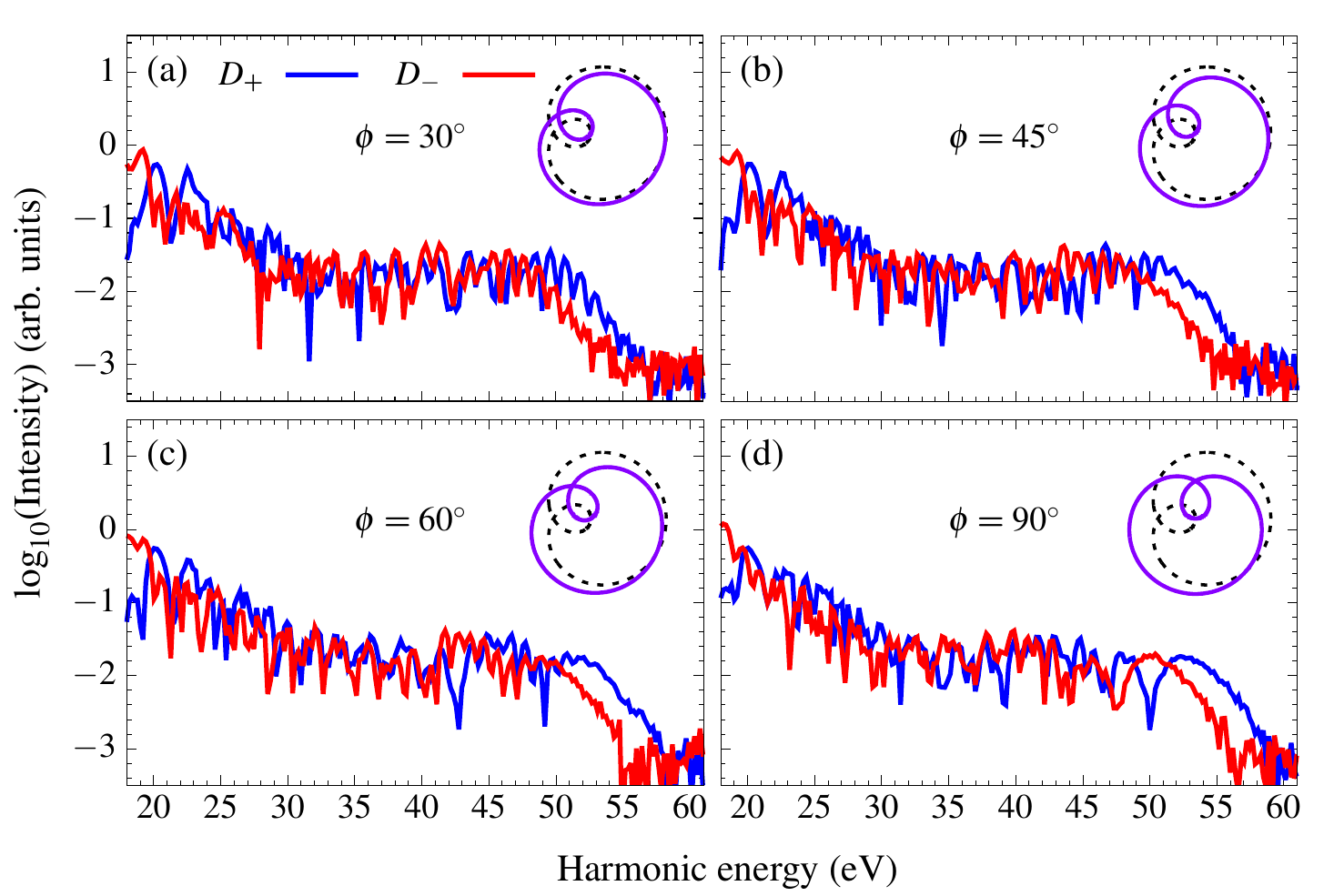}
	\caption{Sensitivity of the harmonic spectra with respect to the sub-cycle phase, $\phi$, between  
		$\omega-2\omega$ fields. 
		(a) $30^\circ$, (b) $45^\circ$, (c) $60^\circ$, and (d) $90^\circ$. 
		Insets show Lissajous figures corresponding to co-rotating driving fields (solid purple line) 
		with the Lissajous figure for $\phi = 0^\circ$ in a black dotted line for comparison purposes.
		$\lambda_{1}$ = 1200 nm, and $I_{1} = I_{2} = 5\times 10^{13}$ W/cm$^2$ are used to simulate the spectra.}
	\label{hhg1200_cep}
\end{figure}

Figure \ref{hhg1200_cep} presents harmonic spectra for different values of $\phi$ for  1200 nm 
wavelength of  $\omega$-field with  1:1 intensity ratio. 
As the value of $\phi$  is tuned from $\phi = 0^{\circ}$ to $\phi = 30^{\circ}$, 
spectrum shown in Fig. \ref{hhg1200_cep}(a) displays high contrast of  the $D_{+}$ harmonic component,
co-rotating with the driving field, compared to the counter-rotating $D_{-}$ harmonic component 
in the cutoff region.
The insensitivity of the features of the harmonic spectra can be understood by analysing  the Lissajous figure.
The orientation of the Lissajous figure rotates by changing  $\phi$ value, but 
dynamical rotational symmetry is still absent, as visible from the inset. 
As a result, the polarization and intensity of the emitted harmonics remain unaffected for different values 
of $\phi$, as evident from the spectra shown in Figs. \ref{hhg1200_cep}(b) - \ref{hhg1200_cep}(d).

\begin{figure}[t]
	\centering\includegraphics[width=1\columnwidth]{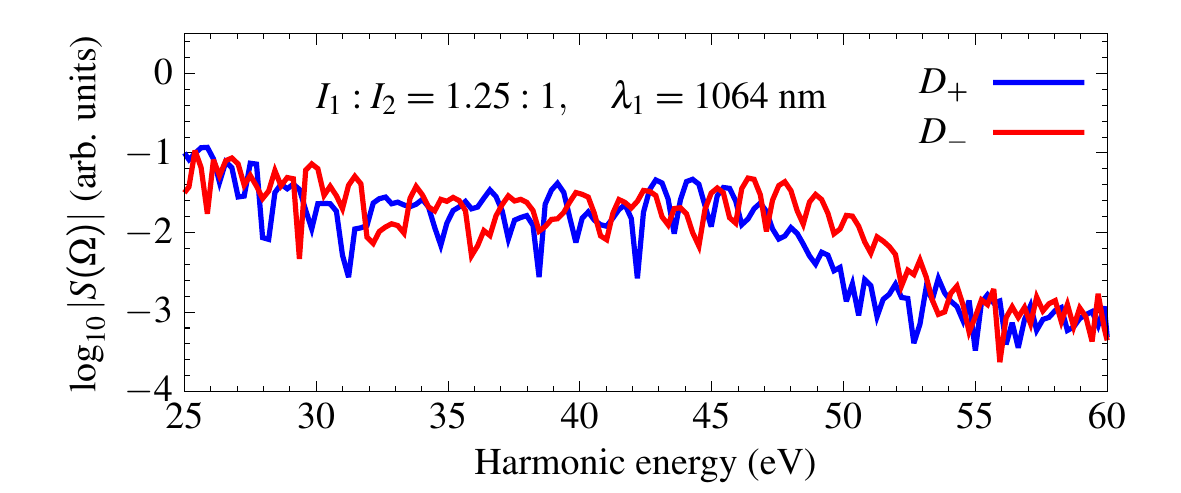}
	\caption{Same as Fig. \ref{hhg_1064_1200}(b) except the rotation direction of combining fields is reversed. This implies that the $D_-$ harmonic component represented by the red line is now co-rotating with the driving field, while the blue line representing the counter-rotating $D_+$ harmonic component.}
	\label{hhg_reverse}
\end{figure}

We have also simulated the harmonic spectra when the helicity of the co-rotating driving fields  is reversed. In this case,  the co-rotating harmonic component dominates in the cutoff region regardless of the rotation direction of the driving field as can be seen in Fig. \ref{hhg_reverse}. 
This offers the possibility of generating attosecond pulses with desired handedness by simply changing the rotation direction of the driving laser pulses.

From the analysis of Figs. \ref{1064_laser_hhg} - \ref{hhg1200_cep}, 
it is established that the essential features of the harmonic spectra, 
such as  domination of the co-rotating $D_+$ component, are robust with respect to the variations in the laser parameters. 
This eliminates the need for precise adjustments of the intensity ratio or the relative phase between the driving fields or a specific choice of wavelength. 
Thus, co-rotating $\omega-2\omega$ scheme can be utilized to 
synthesize attosecond pulse with controlled polarization.

\begin{figure*}[t]
	\centering\includegraphics[width=1.75\columnwidth]{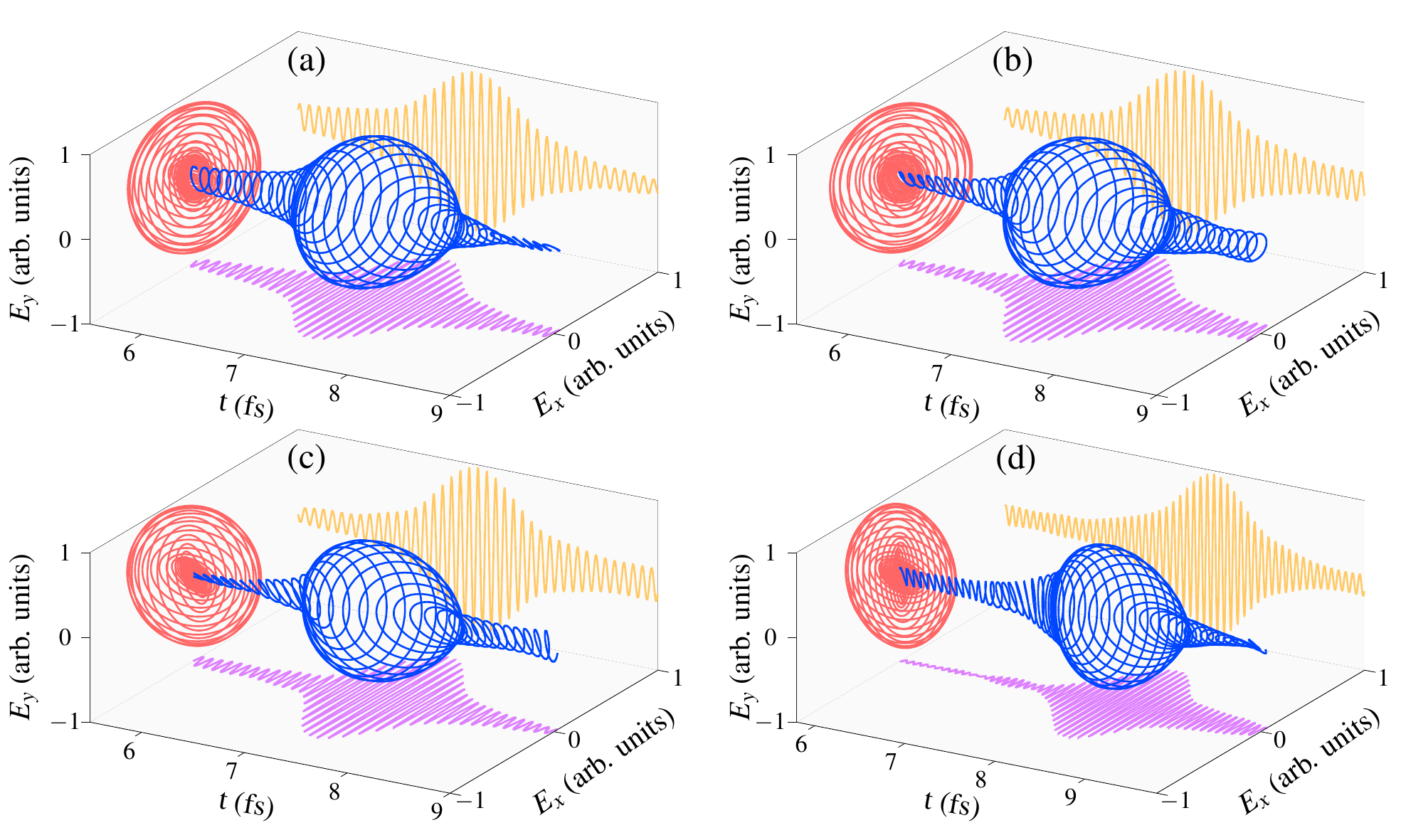}
	\caption{Temporal profile of the synthesized attosecond pulse (blue line) with its 
		$x$ component (magenta line), $y$ component (yellow line), and Lissajous figure (red line) of the total electric field. Different attosecond pulses 
		are generated by superposing the harmonics near the cutoff region of the harmonic  spectra presented in the following figures: (a) Fig. \ref{1064_laser_hhg}, (b) Fig. \ref{hhg_1064_1200}(a), (c) Fig. \ref{hhg_1064_1200}(b), and  (d) Fig. \ref{hhg_1064_1200}(d). 
		The synthesized  pulses  have 
		pulse duration $\sim 550 - 600$ attoseconds, and  ellipticity $\sim  0.77 - 0.88$. }
	\label{fig_asp}
\end{figure*}

\begin{figure}[t]
	\centering\includegraphics[width=1\columnwidth]{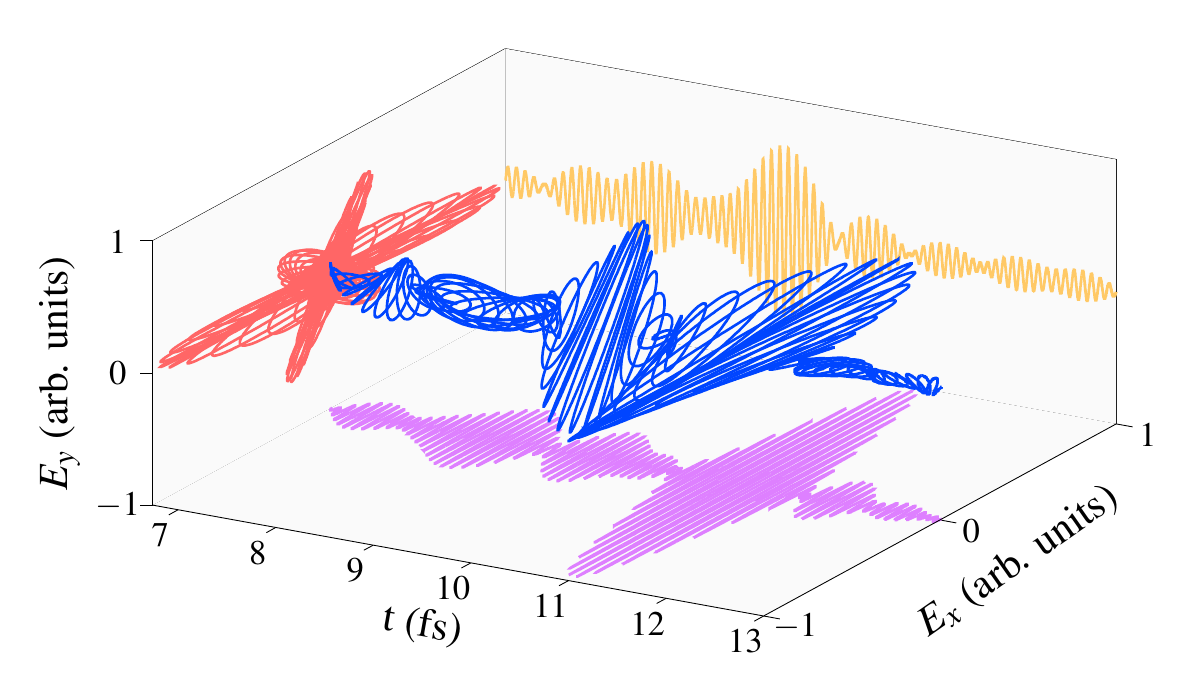}
	\caption{Temporal profile of synthesized attosecond pulses using counter-rotating $\omega-2\omega$ driving fields. All other parameters are similar to those used in Fig. \ref{fig_asp}(a). 
		The synthesized attosecond pulses have ellipticity $\sim  0.13$.}
	\label{fig_ASPcounter}
\end{figure}

To illustrate  the feasibility  of generating attosecond pulses with tunable polarization via HHG driven by 
co-rotating $\omega-2\omega$ fields configuration, let us focus on the harmonic spectra  
in the energy range $44 - 51$ eV.
Figure \ref{fig_asp} shows the temporal profile of the synthesized attosecond pulse with its 
$x$-component, $y$-component and the Lissajous figure of the total electric field. 
The pulse shown in Fig. \ref{fig_asp}(a) has ellipticity as high as $0.88$ with $\sim 630$ attoseconds pulse duration. This elliptical pulse is synthesized by superposing the 
harmonics in the energy range $44-51$ eV from the spectra shown in Fig. \ref{1064_laser_hhg}.  
The high ellipticity is a direct consequence of unequal intensities of the two co- and counter-rotating  harmonic components, i.e., $D_{+}$ and $D_{-}$.  
In contrast to the co-rotating driving fields, if one considers the counter-rotating driving fields to generate high-harmonics, the resultant attosecond pulse exhibits ellipticity  as low as $\sim  0.13$ [see Fig. ~\ref{fig_ASPcounter}]. Thus, co-rotating $\omega-2\omega$  bicircular fields is a potential way to generate 
highly elliptical attosecond pulses. 

To demonstrate  the robustness and universality,  
we have also synthesized pulses from the spectra shown in 
Figs. \ref{hhg_1064_1200}(a),  \ref{hhg_1064_1200}(b), and  \ref{hhg_1064_1200}(d); and 
the corresponding attosecond pulses  are displayed in Figs. \ref{fig_asp}(b) - \ref{fig_asp}(d), respectively. 
The filtered harmonic window is considered near the cutoff region of the respective harmonic spectrum. 
In all cases, the range  of the pulse duration and the ellipticity of  the synthesized  pulses are 
$\sim 550 - 600$ attoseconds and $\sim  0.77 - 0.88$, respectively.  
As expected from the harmonic spectra, 
the helicity of the generated attosecond pulses is same as the driving laser field. 
The high ellipticity of the synthesized attosecond pulses proves the potential 
of the co-rotating $\omega-2\omega$ fields scheme. 
Moreover, the absence of rotational symmetry in the co-rotating field configuration 
translates into the robustness of the generated HHG spectra, wherein small changes in the parameters of the driving fields leave the spectra unaltered.
 
\section{Summary and Conclusions}
\label{sec4}
In summary, we have successfully demonstrated  the generation of  
attosecond pulses with tunable ellipticity via 
HHG driven by co-rotating $\omega-2\omega$ bicircular fields.  
The absence of  the 
dynamical rotational symmetry in the co-rotating $\omega-2\omega$  fields 
translates into the generation of the harmonics with same helicity, which leads to elliptically polarised  
attosecond pulses. 
It is found that the  essential features of the generated harmonics, via co-rotating $\omega-2\omega$ fields configuration, remain insensitive against variations in laser parameters, such as fundamental driving wavelength, intensity ratio, and the subcycle phase between two fields. 
Moreover, the effect of the focal averaging is expected to negligible as the attosecond pulses are synthesised 
using harmonics in near-cutoff region ~\cite{Li2022_PRA}.  
Thus, it avoids the need for precise adjustments of laser parameters from the experimental perspective.
Moreover, the reliance of the polarization properties of the 
harmonics  on driving fields' parameters provides opportunities 
to shape the polarization  of the generated attosecond pulses. 
The generated chiral attosecond pulses can be employed  to study chiral-sensitive dynamics on its intrinsic timescales \cite{Fan2015_PNAS, Cireasa2015_NatPhys, Tang2011_Scienece, Hendry2010_NatuteNanotech, Baykusheva2018_PRX}.
Furthermore, our work can extend to various scenarios, such as 
current-carrying molecular states, which can further increase the ellipticity of the pulse~\cite{Xie2008_PRL, Zhang2020_PRA} or different pulse shaping techniques to extend the harmonic cutoffs~\cite{Holkundkar_2020, PhysRevLett.117.093003, PhysRevA.96.033407}. 

\section*{Acknowledgments} A. R. H. acknowledges support from Science and Engineering Research Board (SERB) India 
(CRG/2020/001020). G. D. acknowledges support from SERB India 
(Project No. MTR/2021/000138). 
  

%

\end{document}